\documentclass[aps,prl,twocolumn,superscriptaddress,showpacs,wasysym,floatfix]{revtex4-1}
\usepackage{bm,amssymb,amsmath,graphicx}
\begin{document}
\title{Classical and quantum magneto-oscillations of current flow near a p-n junction in graphene}

\author{Aavishkar A. Patel}
\affiliation{Physics Department, Lancaster University, Lancaster, LA1 4YB, UK}
\affiliation{Department of Physics, Indian Institute of Technology Kanpur, Kanpur 208016, India}

\author{Nathan Davies}
\affiliation{Physics Department, Lancaster University, Lancaster, LA1 4YB, UK}

\author{Vadim Cheianov}
\affiliation{Physics Department, Lancaster University, Lancaster, LA1 4YB, UK}

\author{Vladimir I. Fal'ko}
\affiliation{Physics Department, Lancaster University, Lancaster, LA1 4YB, UK}


\begin{abstract}
The proposed semiclassical theory predicts two types of oscillations in the flow of current injected from a point source near a ballistic p-n junction in graphene in a strong magnetic field. One originates from the classical effect of bunching of cyclotron orbits of electrons passing back and forth across the p-n interface, which displays a pronounced dependence on the commensurability between the cyclotron radii in the n- and p-regions. The other effect is caused by the interference of monochromatic electron waves in p-n junctions with equal carrier densities on the two sides and it consists in magneto-oscillations in the current transmission through the interface with periodicity similar to Shubnikov-de Haas oscillations.
\end{abstract}
\maketitle
Graphene is a gapless semiconductor with charge carriers that behave like massless Dirac particles~\cite{Review,2DEG}, in which it is possible to locally control the carrier density and type using local electrostatic gates, and create p-n junctions~\cite{Exp1,Exp2,Exp3,Exp4}. Due to the Dirac-type properties of the carriers, such p-n junctions are highly transparent for incoming electrons~\cite{GraphenePNJ,Tunneling}. Also, it has been suggested that p-n junctions in ballistic graphene may be able to focus flow of electrons injected through a point-like source~\cite{Veselago}, which is the result of the inverted dispersions $\epsilon_c=vp$ and $\epsilon_v=U-vp$, for electrons in the conduction and valence bands on opposite sides of the junction, and consequently, an effectively negative index of refraction for electron trajectories.
\par
In this paper we study the features in the flow of electrons injected in the vicinity of a p-n junction in ballistic graphene in a strong magnetic field, resulting from the negative refraction of electrons crossing the p-n interface. Reflected and transmitted electrons follow trajectories which combine elements of skipping orbits~\cite{Bohr,Teller,Halperin} (formed by repeatedly reflected electrons) and snaking orbits (formed by electrons repeatedly crossing the p-n interface). In a recent experimental work~\cite{Exp5}, signatures of such snaking orbits were observed in transport measurements on gated graphene p-n junctions. Also, It has been noticed~\cite{Beenakker,Previous} that when skipping orbits originate from a point-like source, they bunch into caustics and exhibit focusing of charge flow at periodically repeated cusps. Here, we show that periodic appearance of caustics and their cusps is also characteristic for the combined skipping-snaking orbits of electrons propagating along a p-n interface in ballistic graphene, and we analyse how the resulting singularities in the flow of current injected using a point-like source depend on the relative value of the densities of carriers (electrons/holes) on the two sides of the p-n junction.
\begin{figure}
\centering
\includegraphics[width=8.5cm]{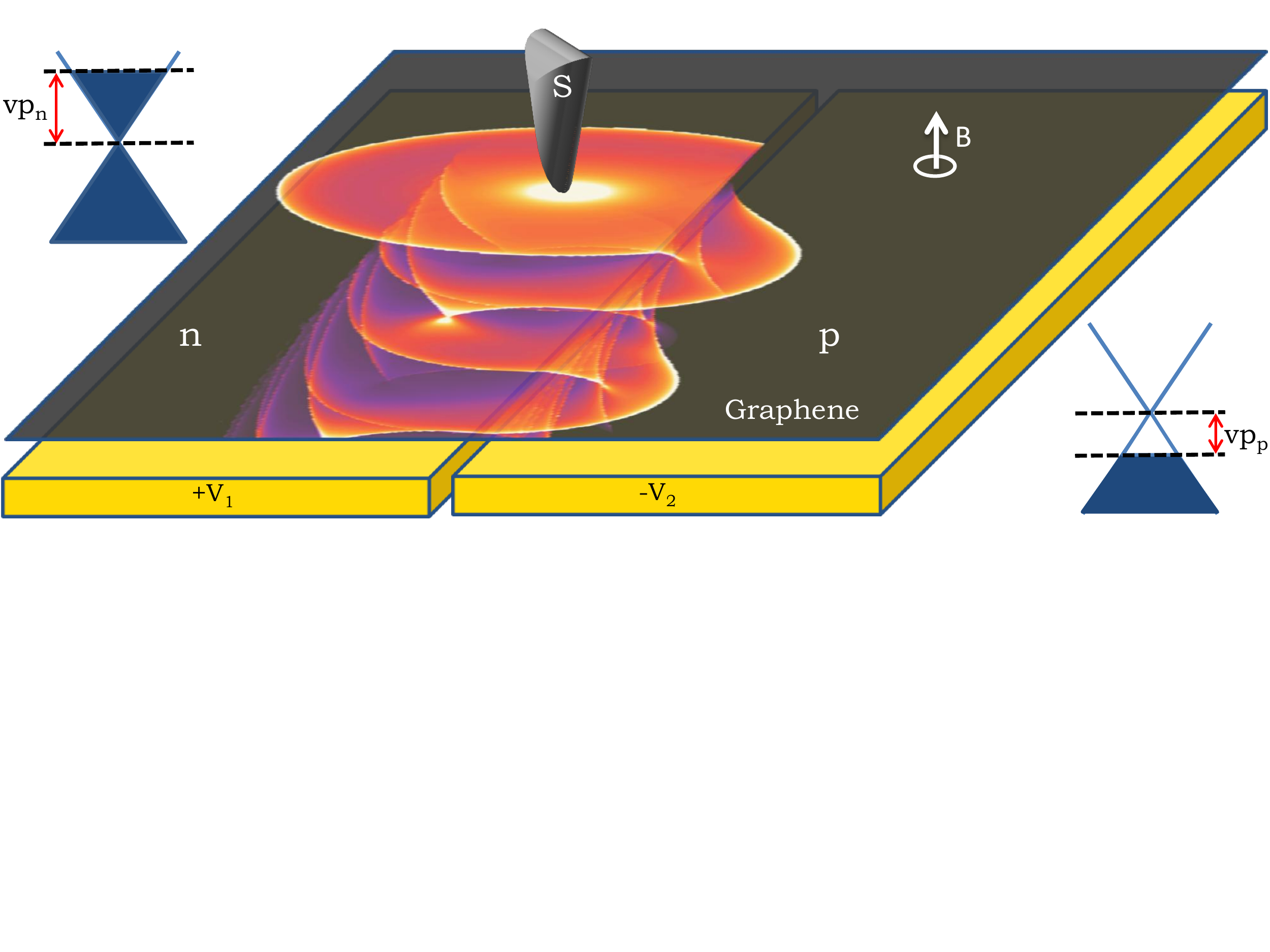}
\vspace{-3cm}
\caption{Current distribution in graphene p-n junction in a perpendicular magnetic field, with electrons injected isotropically at the Fermi energy from the point source S.}
\label{Fig:device}
\end{figure} 
A typical distribution of current calculated in such a system is illustrated in Fig.~\ref{Fig:device}, where the formation of the p-n junction is defined using a pair of split gates providing a potential step $U=v(p_n+p_p)$ for electrons in graphene, and the maxima in the current intensity are marked as the bright spots. Also, we find that for a junction with equal carrier densities on the two sides, the distribution of current displays additional quantum magneto-oscillations.
\par
\begin{figure}
\centering
\includegraphics[width=8.5cm]{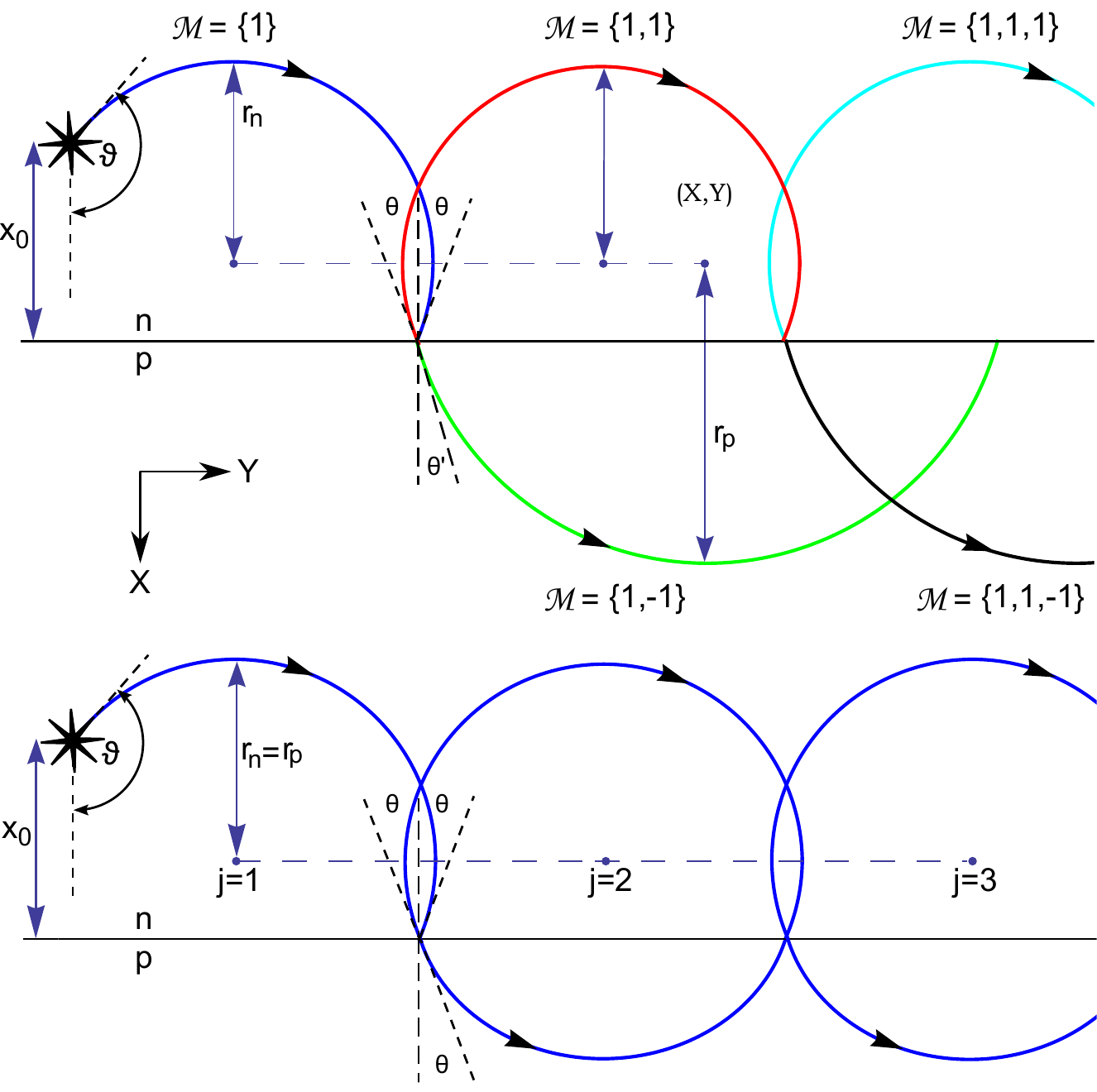}
\caption{Branching of trajectories of electrons propagating in a magnetic field along an asymmetric, $r_n \neq r_p$ (top), and symmetric, $r_n = r_p$ (bottom) p-n junction.}
\label{Fig:paths}
\end{figure}

The proposed theory is based on the analysis of the families of orbits of electrons injected, \textit{e.g.}, on the n-side of the p-n junction at a distance $x_0$ from it. The radii of the electron cyclotron orbits on either side of the interface are given by $r_{n(p)}=p_{n(p)}/eB$, where $p_{n(p)}$ are the electron Fermi momenta in the doped regions. Also, as in Ref.~\cite{Tunneling}, we assume the p-n junction potential step $U$ to be sharp on the scale of the electron Fermi wavelength. For each individual trajectory, such as shown in Fig.~\ref{Fig:paths}, an electron leaves the source at a certain angle $\vartheta$, $0<\vartheta<2\pi$. The following path of the electron consists of a sequence of circular segments on either the n-doped (with radius $r_n$) or the p-doped (with $r_p$) side, which are matched at the interface by the kinematically prescribed Snell's law with negative refraction index~\cite{Veselago},
\begin{equation}
r_{n}\sin\theta = -r_{p}\sin\theta^{\prime},~\sin\theta = -\sin\vartheta - \frac{x_0}{r_n}.
\label{Eq:Snell}
\end{equation}
Here, the values of electron momenta $p_{n(p)}$ (where $v(p_n+p_p)=U$) are absorbed in the values of the cyclotron radii $r_{n(p)}$, and $\theta$ is the angle of incidence and specular reflection, while $\theta^{\prime}$ is the angle of refraction. The sign of the refraction index in Eq.(\ref{Eq:Snell}) is negative because the group velocity is $\vec{v}_c=v\vec{p}/p$ in the conduction band and $\vec{v}_v=-v\vec{p}/p$ in the valence band. The latter relation also prescribes the reversal of the direction of the electron's angular velocity as it crosses the interface, making electrons drift along the p-n interface in a magnetic field (Fig.~\ref{Fig:paths}).
\par
To describe the distribution of the current carried by the drifting electrons, one has to take into account that electron trajectories branch at the interface, with the probability $W_T(\theta)$ to be transmitted and $W_R=1-W_T$ to be reflected. For electrons arriving at a sharp p-n interface with angle of incidence $|\theta|<\theta_c$~\cite{Tunneling,GraphenePNJ},
\begin{equation}
W_{T}=\frac{4 \cos \theta \sqrt{1-\kappa^{2} \sin^{2} \theta}}{(\cos \theta + \sqrt{1-\kappa^{2} \sin^{2} \theta})^{2} + (1 + \kappa)^2 \sin^2 \theta},
\end{equation}
and $W_T=0$ for $|\theta|\geq\theta_c$, where $\sin \theta_c=1/\kappa$ and $\kappa=r_n/r_p$. Branching of the electron trajectories can be then labeled by a sequence,
\begin{equation}
\mathcal{M}=\{s_{1},s_{2},.....,s_{N(\mathcal{M})-1},s_{N(\mathcal{M})}\},
\end{equation}
where $s_i=+1$ if the $i$\textsuperscript{th} circle segment is on the n-side and $s_i=-1$ if it is on the p-side of the junction. Together with $\vartheta$, the sequence uniquely defines the path with $N$ circle segments, as shown in Fig.~\ref{Fig:paths}. Using Snell's law~(\ref{Eq:Snell}) and elementary geometry, one can show that the centers of all cyclotron orbits whose segments belong to the sequence $\mathcal{M}$ are positioned at the same distance $X(\vartheta)$ from the p-n interface.
\begin{figure*}
\centering
\includegraphics[width=\textwidth]{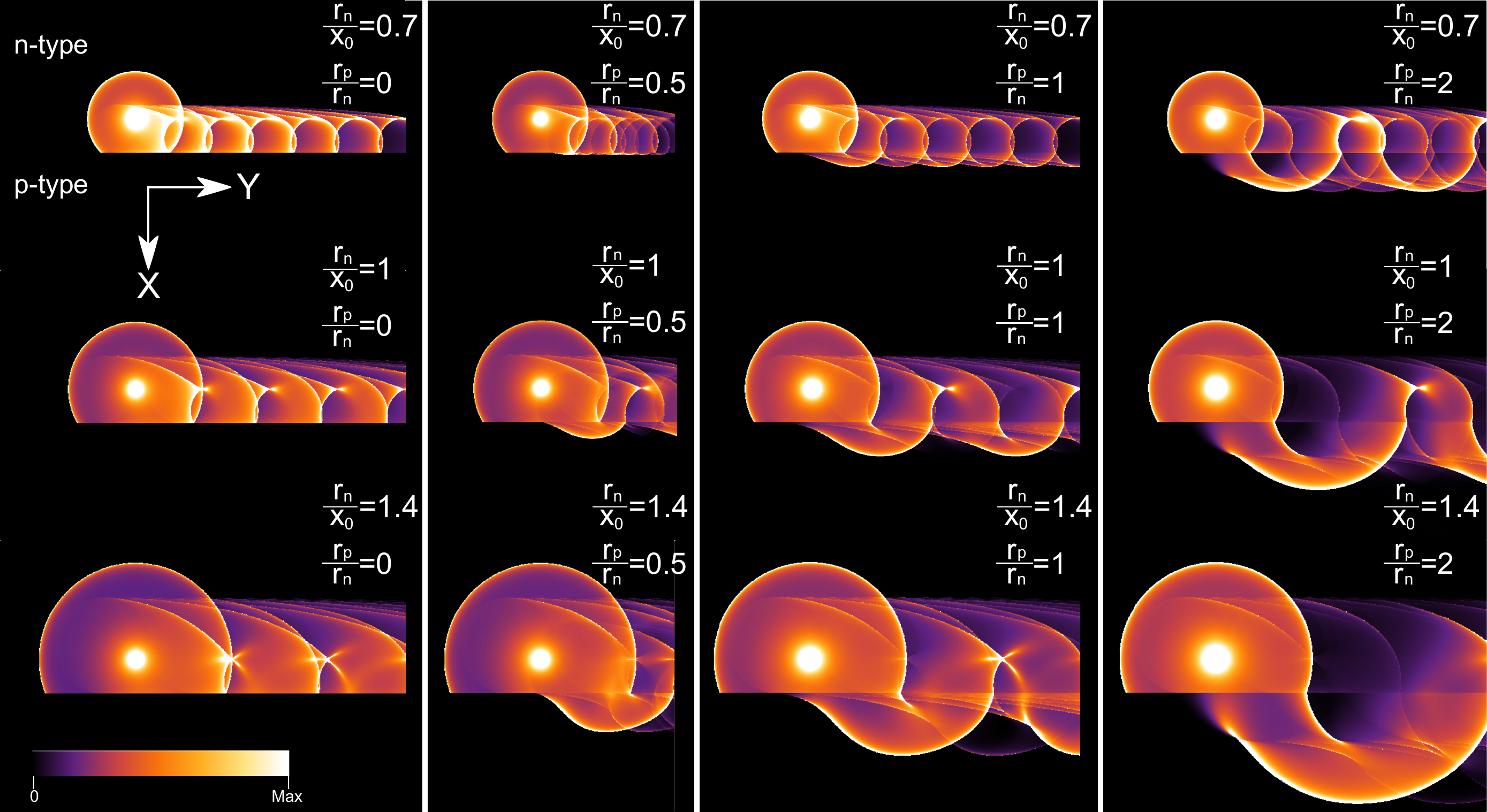}
\caption{Calculated density of current injected from a point-like source (bright spot) near the p-n junction. Electrons are injected isotropically at a distance $x_0$ from the interface and travel along the interface undergoing transmissions and reflections. Spatial oscillations of current distribution are caused by a periodic appearance of caustics and cusp singularities}
\label{Fig:patterns}
\end{figure*}
\par
To calculate the density $\varrho(x,y)$ of trajectories on the graphene sheet, we define a generating function~\cite{Previous},
\begin{widetext}
\begin{eqnarray}
&&F(x,y,\vartheta,\mathcal{M}) \equiv \left[x-X(\vartheta)\right]^{2} + \left[y-Y(\vartheta,\mathcal{M})\right]^{2} - r^2(\mathcal{M}), \nonumber \\
&&Y(\vartheta,\mathcal{M}) = \sum_{i=1}^{N(\mathcal{M})-1} \left[\left(z_{i}(\mathcal{M})+2\right) \frac{\sqrt{r_n^{2} - X^2(\vartheta)}}{2} - \left(z_{i}(\mathcal{M})-2\right) \frac{\sqrt{r_p^{2} - X^2(\vartheta)}}{2}\right] - r_n \cos \vartheta, \nonumber \\
&&X(\vartheta) = x_{0} + r_n \sin \vartheta,~r(\mathcal{M}) = s_{N(\mathcal{M})}\frac{r_n-r_p}{2} + \frac{r_n+r_p}{2},~z_{i}(\mathcal{M}) = s_{i}(\mathcal{M}) + s_{i+1}(\mathcal{M}).
\end{eqnarray}
For each $\vartheta$ and $\mathcal{M}$, a point $(x,y)$ lies on the $N$\textsuperscript{th} segment of the path if $F=0$, thus, the density of trajectories is
\begin{eqnarray}
&&\varrho(x,y)=\sum_{\mathcal{M}}\left[1-s_{N(\mathcal{M})}\mathrm{sign}x\right]\left|\frac{\partial F}{\partial \vartheta}\right|^{-1}_{F=0}(W_T)^{n_T(\mathcal{M})}(1-W_T)^{n_R(\mathcal{M})}, \\
&&n_{R}(\mathcal{M})=\frac{N(\mathcal{M})+\tilde{S}(\mathcal{M})-1}{2},~n_{T}(\mathcal{M})= \frac{N(\mathcal{M})-\tilde{S}(\mathcal{M})-1}{2},~\tilde{S}(\mathcal{M})=\sum_{i=1}^{N(\mathcal{M})-1} s_{i}(\mathcal{M}) s_{i+1}(\mathcal{M}), \nonumber
\label{Eq:cli}
\end{eqnarray}
\end{widetext}
where we take into account all sequences whose final segments pass through $(x,y)$ for all possible $\vartheta$ which solve equation $F=0$, and $n_R$($n_T$) is the number of reflections(transmissions) contained in the sequence $\mathcal{M}$.
\par
Figure~\ref{Fig:patterns} shows the spatial distribution of $\varrho(x,y)$, calculated from Eq.~(\ref{Eq:cli}), which illustrates spreading of current flow injected from a point-like source along the p-n interface, for various combintations of system parameters: $x_0$, $r_n$ and $r_p$. It reflects the periodic appearance of singularities in the skipping-snaking orbits (caustics and caustic cusps), weighted with the reflection/transmission probabilities. In all panels in Fig.~\ref{Fig:patterns}, caustics are seen as bright lines, indicating the points $(x_c,y_c)$ on the 2D plane where both $F=0$ and $\partial F/\partial\vartheta=0$:
\begin{eqnarray}
&&x_{c}(\vartheta,\mathcal{M})=X(\vartheta) \pm \frac{\beta(\vartheta,\mathcal{M})r(\mathcal{M})}{\sqrt{\alpha^2(\vartheta)+\beta^2(\vartheta,\mathcal{M})}}, \\ 
&&y_{c}(\vartheta,\mathcal{M})=Y(\vartheta,\mathcal{M}) \pm \frac{\alpha(\vartheta)r(\mathcal{M})}{\sqrt{\alpha^2(\vartheta)+\beta^2(\vartheta,\mathcal{M})}}, \nonumber \\
&&\alpha(\vartheta)=dX(\vartheta)/d\vartheta,~\beta(\vartheta,\mathcal{M})=dY(\vartheta,\mathcal{M})/d\vartheta. \nonumber
\end{eqnarray}
Cusps, representing the bright spots of \textquotedblleft magnetic focusing\textquotedblright of the electron flow~\cite{Beenakker,Previous}, appear at the points where, additionally, $\partial^2 F/\partial^2 \vartheta=0$.
\begin{figure}
\centering
\includegraphics[width=8.5cm]{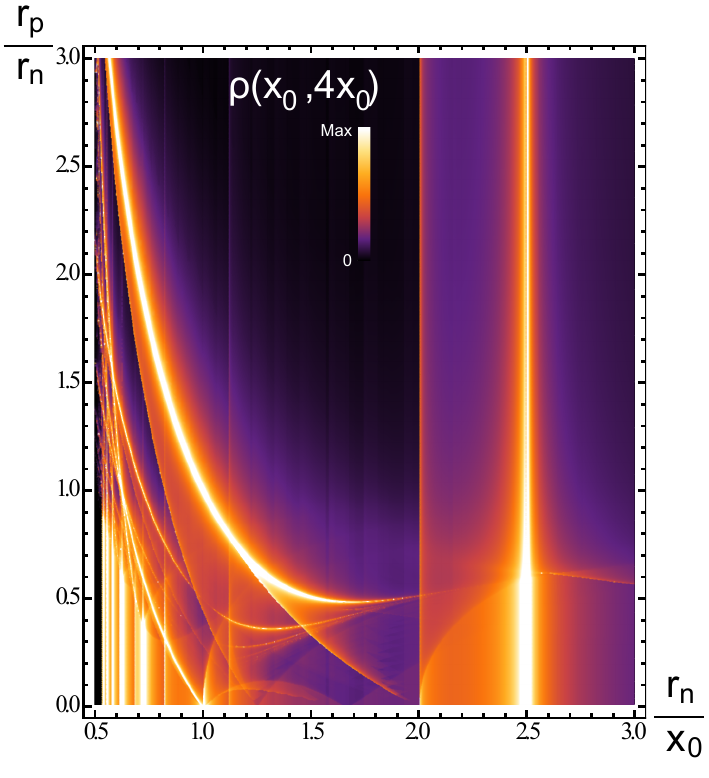}
\caption{Calculated local density of current at the point $(x_0,4x_0)$, located directly below the source at a distance of $4x_0$ from it. The magnetic field $B$ and carrier density $n_{n(p)}$ dependence of the local current density are encoded in the dependence of $r_p/r_n=\sqrt{n_p/n_n}$ and $r_n/x_0\propto1/B$.}
\label{Fig:square}
\end{figure}  
\par
When $r_p=0$ (1\textsuperscript{st} column of Fig.~\ref{Fig:patterns}), all electrons striking the interface on the n-side are reflected, so that, classically, current would flow only on the side of the source, displaying singularities (single cusps as well as cusp triplets) in the distribution specific for the families of simple skipping orbits~\cite{Previous}. As $r_p$ is increased (2\textsuperscript{nd} column), electrons start getting transmitted to the p-side, leading to caustics and cusp-like features on both sides of the interface. As $r_p$ is made comparable to $r_n$, the critical angle $\theta_c$ disappears, and the strength of features produced by reflected electrons diminishes significantly (3\textsuperscript{rd} column). In fact, for $x_0=r_n=r_p$, the caustics and cusps produced by skipping trajectories vanish, and we only find the features produced by snaking orbits, leading to a doubling of the period of appearance of cusps, as compared to the case of purely skipping trajectories. As $r_p$ is increased further (4\textsuperscript{th} column), the period of appearance of cusps of snaking orbits and the number of cusps of skipping orbits between consecutive cusps of snaking orbits increase. Note that, in the general case of $r_n \neq r_p$, the trajectories for different paths are not commensurate for all angles. This determines a complex pattern of current distribution. In special cases where the radii $r_n$ and $r_p$ are commensurate ($r_p=mr_n$), the singularities in the current distribution on the side of the source or the opposite side of the p-n junction display long range periodicity, at $mr_n$. The evolution of the current distribution upon the change of magnetic field and carrier density is shown in Fig.~\ref{Fig:square}, for current measured at a distance of $4x_0$ from the source.
\par
In the special case of a symmetric p-n junction, with $r_{n}=r_{p}$, it is also possible to develop a coherent semiclassical description of current distribution, taking into account quantum interference effects. This involves taking into account the phases acquired by electrons upon transmission/reflection at the interface and during free propagation of Dirac electrons. As shown in Fig.~\ref{Fig:paths}, for the case $r_n=r_p$, all families $\mathcal{M}$ of electron trajectories for any $x_0$ and $\vartheta$ belong to a periodic sequence of circles, which enables one to describe the wave propagating along such paths using a pair of amplitudes, $\psi_n(j,\vartheta)$ and $\psi_p(j,\vartheta)$ for an electron on the n- or p-side of the interface after $j-1$ encounters with it. The wave amplitudes $\psi_{n(p)}(j,\vartheta)$ on the $j$\textsuperscript{th} circle segment on the n(p) side, immediately after the $(j-1)$\textsuperscript{th} interaction with the interface, were found as, 
\begin{eqnarray}
&\left(
\begin{array}{clrr} 
     \psi_{n}(j,\vartheta)  \\ 
     \psi_{p}(j,\vartheta)
\end{array}
\right)=\left[\Xi S\right]^{j-2}\Xi
\left(
\begin{array}{clrr} 
     e^{i\phi_0}  \\ 
     0
\end{array}
\right), \label{Eq:amplitudes} \\
\Xi=e^{i\theta}&\left(
\begin{array}{clrr} 
    i\sin{\theta} & \cos{\theta} \\ 
    \cos{\theta}  & i\sin{\theta}
\end{array}
\right),~
S=\left(
\begin{array}{clrr} 
    e^{i\phi_n} & 0  \\ 
    0 & e^{i\phi_p}
\end{array}
\right). \nonumber
\end{eqnarray}
Here the scattering matrix $\Xi$ describes the transmission and reflection of Dirac spinors at the interface, and $S$ their propagation between interactions with the interface. These matrices were calculated for the plane wave Dirac spinors of electrons in graphene~\cite{GraphenePNJ,Tunneling}, and hence the product $S\Xi$ includes the electron Berry phase. The phase $\phi_0$ is acquired by electrons while propagating between the source and the interface, and $\phi_{n(p)}=\int(-e\vec{A}/\hbar+\vec{p}_{n(p)}/\hbar)\cdot d\vec{l}$ are phases acquired by electrons propagating in a magnetic field along a single circular segment. As segments on the p- and n-sides always form full circles, Fig.~\ref{Fig:paths}, we find that $\phi_n-\phi_p=\pi p^2/(\hbar eB)$.
\par
Due to the above-discussed feature, the drift of the Dirac electron along the p-n interface described by Eq.~(\ref{Eq:amplitudes}) has the following interesting property: for any $\vartheta$ and $x_0$, both the matrix $V=\Xi S\Xi$ and the matrix $W=\left[S\Xi\right]^2$ are diagonal. According to Eq.~(\ref{Eq:amplitudes}), after $2N$ encounters with the interface, the amplitudes $\psi_{n(p)}(2N+1,\vartheta)$ of the Dirac electron on the n(p)-sides are determined by the evolution matrix $(\Xi S)^{2N-1}\Xi=VW^{N-1}$. Therefore, if $p^2/(\hbar eB)=(2l+1)$, the electrons will appear only on the same side of the junction as their source after any even number of encounters with the p-n interface. As a result, we predict rapid quantum oscillations of the electron current between the two sides of the junction. These oscillations develop on a scale linear in the inverse magnetic field $B^{-1}$, with the same period $\Delta(B^{-1})=2\hbar e/p^2$ as Shubnikov-de Haas oscillations~\cite{2DEG}: the phase difference, $\phi_n-\phi_p$, coincides exactly with the phase acquired by an electron with momentum $p$ after it has propagated along a closed cyclotron trajectory in the bulk of the graphene flake with carrier density $n=p^2/(\pi\hbar^2)$. 
\par
At a finite temperature $T\gtrsim\hbar v/(4k_Br)$, $r\equiv r_n=r_p$, thermal smearing of the Fermi level results in the loss of monochromaticity of the electron source, which attenuates the above-mentioned intereference effect, leading to the classical results applicable to the incoherent propagation of electrons. Then, incoherent propogation gives rise to a classical density profile for the distribution of electrons injected by the DC source, as defined in Eq.~(\ref{Eq:cli}). For the case of $r_n=r_p$, summation of the series in Eq.~(\ref{Eq:cli}) can be performed using a recursive method similar to Eq.~(\ref{Eq:amplitudes}), resulting in 
\begin{equation}\nonumber
\varrho(x,y)=\sum_{j} \frac{1}{2}\left[1-\mathrm{sign}x(1-2W_{T}[\theta])^{j-1}\right]\left|\frac{\partial F}{\partial \vartheta}\right|^{-1}_{F=0}, 
\end{equation}
where $\vartheta$ is chosen to solve $F=0$ for the point $(x,y)$, and which produces an alternating pattern of weak and strong cusps, as shown in Fig.~\ref{Fig:patterns}.
\par
In conclusion, we predict periodic spatial modulation of current injected from a point source near a p-n junction in graphene in a magnetic field, which originates from caustic bunching of skipping-snaking orbits. Experimentally, such oscillations would appear as magneto-oscillations of conductance of a two-terminal device with two point contacts. We also find that, for the case of commensurate size of electron cyclotron orbits in the n-type and p-type regions, the current distribution undergoes rapid quantum oscillations between the two sides.
\par
The authors acknowledge support from the ERC Advanced Investigator grant \textquotedblleft Graphene and beyond\textquotedblright and the Royal Society.
\bibliographystyle{apsrev1}
\bibliography{paper}  

\begin{thebibliography}{15}
\expandafter\ifx\csname natexlab\endcsname\relax\def\natexlab#1{#1}\fi
\expandafter\ifx\csname bibnamefont\endcsname\relax
  \def\bibnamefont#1{#1}\fi
\expandafter\ifx\csname bibfnamefont\endcsname\relax
  \def\bibfnamefont#1{#1}\fi
\expandafter\ifx\csname citenamefont\endcsname\relax
  \def\citenamefont#1{#1}\fi
\expandafter\ifx\csname url\endcsname\relax
  \def\url#1{\texttt{#1}}\fi
\expandafter\ifx\csname urlprefix\endcsname\relax\def\urlprefix{URL }\fi
\providecommand{\bibinfo}[2]{#2}
\providecommand{\eprint}[2][]{\url{#2}}

\bibitem[{\citenamefont{Neto et~al.}(2009)\citenamefont{Neto, Guinea, Peres,
  Novoselov, and Geim}}]{Review}
\bibinfo{author}{\bibfnamefont{A.}~\bibnamefont{Neto}},
  \bibinfo{author}{\bibfnamefont{F.}~\bibnamefont{Guinea}},
  \bibinfo{author}{\bibfnamefont{N.}~\bibnamefont{Peres}},
  \bibinfo{author}{\bibfnamefont{K.}~\bibnamefont{Novoselov}},
  \bibnamefont{and} \bibinfo{author}{\bibfnamefont{A.}~\bibnamefont{Geim}},
  \bibinfo{journal}{Reviews of Modern Physics} \textbf{\bibinfo{volume}{81}},
  \bibinfo{pages}{109} (\bibinfo{year}{2009}).

\bibitem[{\citenamefont{Novoselov et~al.}(2005)\citenamefont{Novoselov, Geim,
  Morozov, Jiang, Grigorieva, Dubonos, and Firsov}}]{2DEG}
\bibinfo{author}{\bibfnamefont{K.}~\bibnamefont{Novoselov}},
  \bibinfo{author}{\bibfnamefont{A.}~\bibnamefont{Geim}},
  \bibinfo{author}{\bibfnamefont{S.}~\bibnamefont{Morozov}},
  \bibinfo{author}{\bibfnamefont{D.}~\bibnamefont{Jiang}},
  \bibinfo{author}{\bibfnamefont{M.}~\bibnamefont{Grigorieva}},
  \bibinfo{author}{\bibfnamefont{S.}~\bibnamefont{Dubonos}}, \bibnamefont{and}
  \bibinfo{author}{\bibfnamefont{A.}~\bibnamefont{Firsov}},
  \bibinfo{journal}{Nature} \textbf{\bibinfo{volume}{438}},
  \bibinfo{pages}{197} (\bibinfo{year}{2005}).

\bibitem[{\citenamefont{Huard et~al.}(2007)\citenamefont{Huard, Sulpizio,
  Stander, Todd, Yang, and Goldhaber-Gordon}}]{Exp1}
\bibinfo{author}{\bibfnamefont{B.}~\bibnamefont{Huard}},
  \bibinfo{author}{\bibfnamefont{J.}~\bibnamefont{Sulpizio}},
  \bibinfo{author}{\bibfnamefont{N.}~\bibnamefont{Stander}},
  \bibinfo{author}{\bibfnamefont{K.}~\bibnamefont{Todd}},
  \bibinfo{author}{\bibfnamefont{B.}~\bibnamefont{Yang}}, \bibnamefont{and}
  \bibinfo{author}{\bibfnamefont{D.}~\bibnamefont{Goldhaber-Gordon}},
  \bibinfo{journal}{Physical review letters} \textbf{\bibinfo{volume}{98}},
  \bibinfo{pages}{236803} (\bibinfo{year}{2007}).

\bibitem[{\citenamefont{Williams et~al.}(2007)\citenamefont{Williams, DiCarlo,
  and Marcus}}]{Exp2}
\bibinfo{author}{\bibfnamefont{J.}~\bibnamefont{Williams}},
  \bibinfo{author}{\bibfnamefont{L.}~\bibnamefont{DiCarlo}}, \bibnamefont{and}
  \bibinfo{author}{\bibfnamefont{C.}~\bibnamefont{Marcus}},
  \bibinfo{journal}{Science} \textbf{\bibinfo{volume}{317}},
  \bibinfo{pages}{638} (\bibinfo{year}{2007}).

\bibitem[{\citenamefont{{\"O}zyilmaz et~al.}(2007)\citenamefont{{\"O}zyilmaz,
  Jarillo-Herrero, Efetov, Abanin, Levitov, and Kim}}]{Exp3}
\bibinfo{author}{\bibfnamefont{B.}~\bibnamefont{{\"O}zyilmaz}},
  \bibinfo{author}{\bibfnamefont{P.}~\bibnamefont{Jarillo-Herrero}},
  \bibinfo{author}{\bibfnamefont{D.}~\bibnamefont{Efetov}},
  \bibinfo{author}{\bibfnamefont{D.}~\bibnamefont{Abanin}},
  \bibinfo{author}{\bibfnamefont{L.}~\bibnamefont{Levitov}}, \bibnamefont{and}
  \bibinfo{author}{\bibfnamefont{P.}~\bibnamefont{Kim}},
  \bibinfo{journal}{Physical review letters} \textbf{\bibinfo{volume}{99}},
  \bibinfo{pages}{166804} (\bibinfo{year}{2007}).

\bibitem[{\citenamefont{Gorbachev et~al.}(2008)\citenamefont{Gorbachev,
  Mayorov, Savchenko, Horsell, and Guinea}}]{Exp4}
\bibinfo{author}{\bibfnamefont{R.}~\bibnamefont{Gorbachev}},
  \bibinfo{author}{\bibfnamefont{A.}~\bibnamefont{Mayorov}},
  \bibinfo{author}{\bibfnamefont{A.}~\bibnamefont{Savchenko}},
  \bibinfo{author}{\bibfnamefont{D.}~\bibnamefont{Horsell}}, \bibnamefont{and}
  \bibinfo{author}{\bibfnamefont{F.}~\bibnamefont{Guinea}},
  \bibinfo{journal}{Nano letters} \textbf{\bibinfo{volume}{8}},
  \bibinfo{pages}{1995} (\bibinfo{year}{2008}).

\bibitem[{\citenamefont{Cheianov and Fal’ko}(2006)}]{GraphenePNJ}
\bibinfo{author}{\bibfnamefont{V.~V.} \bibnamefont{Cheianov}} \bibnamefont{and}
  \bibinfo{author}{\bibfnamefont{V.~I.} \bibnamefont{Fal’ko}},
  \bibinfo{journal}{Physical Review B} \textbf{\bibinfo{volume}{74}},
  \bibinfo{pages}{041403} (\bibinfo{year}{2006}).

\bibitem[{\citenamefont{Katsnelson et~al.}(2006)\citenamefont{Katsnelson,
  Novoselov, and Geim}}]{Tunneling}
\bibinfo{author}{\bibfnamefont{M.}~\bibnamefont{Katsnelson}},
  \bibinfo{author}{\bibfnamefont{K.}~\bibnamefont{Novoselov}},
  \bibnamefont{and} \bibinfo{author}{\bibfnamefont{A.}~\bibnamefont{Geim}},
  \bibinfo{journal}{Nature Physics} \textbf{\bibinfo{volume}{2}},
  \bibinfo{pages}{620} (\bibinfo{year}{2006}).

\bibitem[{\citenamefont{Cheianov et~al.}(2007)\citenamefont{Cheianov, Fal'ko,
  and Altshuler}}]{Veselago}
\bibinfo{author}{\bibfnamefont{V.}~\bibnamefont{Cheianov}},
  \bibinfo{author}{\bibfnamefont{V.}~\bibnamefont{Fal'ko}}, \bibnamefont{and}
  \bibinfo{author}{\bibfnamefont{B.}~\bibnamefont{Altshuler}},
  \bibinfo{journal}{Science} \textbf{\bibinfo{volume}{315}},
  \bibinfo{pages}{1252} (\bibinfo{year}{2007}).

\bibitem[{\citenamefont{Bohr}(1972)}]{Bohr}
\bibinfo{author}{\bibfnamefont{N.}~\bibnamefont{Bohr}},
  \emph{\bibinfo{title}{Niels Bohr Collected Works}}, vol.~\bibinfo{volume}{1}
  (\bibinfo{publisher}{Elsevier, Amsterdam}, \bibinfo{year}{1972}).

\bibitem[{\citenamefont{Teller}(1931)}]{Teller}
\bibinfo{author}{\bibfnamefont{E.}~\bibnamefont{Teller}},
  \bibinfo{journal}{Zeitschrift f{\"u}r Physik A Hadrons and Nuclei}
  \textbf{\bibinfo{volume}{67}}, \bibinfo{pages}{311} (\bibinfo{year}{1931}).

\bibitem[{\citenamefont{Halperin}(1982)}]{Halperin}
\bibinfo{author}{\bibfnamefont{B.~I.} \bibnamefont{Halperin}},
  \bibinfo{journal}{Physical Review B} \textbf{\bibinfo{volume}{25}},
  \bibinfo{pages}{2185} (\bibinfo{year}{1982}).

\bibitem[{\citenamefont{Williams and Marcus}(2011)}]{Exp5}
\bibinfo{author}{\bibfnamefont{J.~R.} \bibnamefont{Williams}} \bibnamefont{and}
  \bibinfo{author}{\bibfnamefont{C.~M.} \bibnamefont{Marcus}},
  \bibinfo{journal}{Physical Review Letters} \textbf{\bibinfo{volume}{107}},
  \bibinfo{pages}{046602} (\bibinfo{year}{2011}).

\bibitem[{\citenamefont{Beenakker et~al.}(1988)\citenamefont{Beenakker, Houten,
  and Wees}}]{Beenakker}
\bibinfo{author}{\bibfnamefont{C.}~\bibnamefont{Beenakker}},
  \bibinfo{author}{\bibfnamefont{H.}~\bibnamefont{Houten}}, \bibnamefont{and}
  \bibinfo{author}{\bibfnamefont{B.}~\bibnamefont{Wees}}, \bibinfo{journal}{EPL
  (Europhysics Letters)} \textbf{\bibinfo{volume}{7}}, \bibinfo{pages}{359}
  (\bibinfo{year}{1988}).

\bibitem[{\citenamefont{Davies et~al.}(2012)\citenamefont{Davies, Patel,
  Cortijo, Cheianov, Guinea, and Fal'ko}}]{Previous}
\bibinfo{author}{\bibfnamefont{N.}~\bibnamefont{Davies}},
  \bibinfo{author}{\bibfnamefont{A.~A.} \bibnamefont{Patel}},
  \bibinfo{author}{\bibfnamefont{A.}~\bibnamefont{Cortijo}},
  \bibinfo{author}{\bibfnamefont{V.}~\bibnamefont{Cheianov}},
  \bibinfo{author}{\bibfnamefont{F.}~\bibnamefont{Guinea}}, \bibnamefont{and}
  \bibinfo{author}{\bibfnamefont{V.~I.} \bibnamefont{Fal'ko}},
  \bibinfo{journal}{Phys. Rev. B} \textbf{\bibinfo{volume}{85}},
  \bibinfo{pages}{155433} (\bibinfo{year}{2012}).

\end{thebibliography}
\end{document}